\documentclass[12pt]{article}
\usepackage{amsfonts}
\begin{document}
\title{Aspects of noncommutative descriptions of planar systems in
  high magnetic fields.}  \author{C.~D.~Fosco~\footnote{Electronic
    address: fosco@cab.cnea.gov.ar} and
A.~L{\'o}pez~\footnote{Electronic
    address: lopezana@cab.cnea.gov.ar}
  \\ \\
  {\normalsize\it Centro At{\'o}mico Bariloche - Instituto
Balseiro,}\\
  {\normalsize\it Comisi{\'o}n Nacional de Energ{\'\i}a
At{\'o}mica}\\
  {\normalsize\it 8400 Bariloche, Argentina.}}  \date{\today}
\maketitle
\begin{abstract}
\noindent
We study some aspects of recent proposals to use the noncommutative
Chern-Simons theory as an effective description of some planar
condensed matter models in strong magnetic fields, such as the
Quantum
Hall Effect.  We present an alternative justification for such a
description, which may be extended to other planar systems where a
uniform magnetic field is present.
\end{abstract}
\bigskip \newpage
\section{Introduction}

Noncommutative field theories have recently attracted renewed
attention, mostly because of their relevance for the understanding
of
some phenomena in the context of string theory, like the low energy
limit of open strings in the presence of some special background
field
configurations~\cite{castellani,DN}.

In the condensed matter physics context, non-commutative
Chern-Simons
(NCCS) theories have recently been proposed as effective
descriptions
of the Laughlin states in the Quantum Hall
Effect~\cite{susskind,PP,KS}.  Noncommutative field theories have
also
been used to describe the skyrmionic excitations of the Quantum
Hall
ferromagnet at $\nu=1$~\cite{PA, LMR}.

The physics of a bidimensional system of particles in the presence
of
an external magnetic field has a very rich structure, a phenomenon
which is partly due to the particularities of the Landau level
spectrum for a particle in an external field.  In particular, it is
a
well known fact that when the system is restricted to the lowest
Landau level (LLL), area preserving diffeomorphisms become a
symmetry
of the system~\cite{IK,SS,GCT}.  The restriction to the LLL is
usually
invoked as a consequence of the existence of a large gap between
the
lowest and higher Landau Levels \cite{gj}. However, this
restriction
cannot be defined as a smooth limit of the full (all level) system,
since there is a change in the number of physical degrees of
freedom,
an effect that has been known since the early studies on
Chern-Simons
quantum mechanics~\cite{jackiw}, and entirely analogous to a
similar
reduction from the Maxwell-Chern-Simons theory into the pure
Chern-Simons one~\cite{IK,jackiw}.  The change in the number of
degrees of freedom means that one of the physical variables (a
coordinate, for the particle) is transformed into the canonical
momenta of the remaining variable. Thus the usual invariance under
canonical transformations acquires a much greater relevance, since
it
becomes a spacetime symmetry. As canonical transformations preserve
the phase space volume, the symmetry of the reduced system can be
analogously thought of as invariance under area preserving
diffeomorphisms.

The quantum version of these symmetry transformations necessarily
has
to cope with operator ordering problems, since they involve
canonical
conjugate variables that do not commute in the quantum theory.  In
the
operatorial (canonical) quantization method the use of the Weyl
quantization prescription, is the natural way to introduce the
Moyal
product for phase space functions~\cite{castellani}. Of course, the
same phenomenon can be studied in the path integral framework, for
example by means if the mid-point prescription~\cite{midpoint} to
define the matrix elements of Weyl-ordered products, when they
appear
inside the path integral.

In this work we consider systems described by an action with the
general structure:
\begin{equation}
  \label{eq:defact}
S \;=\; S_m \,+\, S_g \,+\, S_{int} \;,
\end{equation}
where $S_m$ denotes the free action for a system of particles
(either
in its first or second quantized representations), $S_g$ is the
action
for a vector (gauge) field $A_\mu$, and $S_{int}$ corresponds to
the
coupling between the particles and the gauge field. A distinctive
feature of the systems we will analyze is that the free action
$S_m$
will be negligible for the dynamics, due to the presence of a
strong
external magnetic field (defined as part of $S_{int}$). This is
usually stated as the `freezing' of the kinetic energy, and it is a
fundamental requisite for the emergence of a noncommutative
description. We shall argue that the noncommutativity is, for the
kind
of systems we are considering, a property of the {\em
description\/}
used rather than a fundamental symmetry. For the noncommutative
theory
corresponding to a system in the presence of an external magnetic
field there is, as we shall see, also a freedom in the choice of
the
deformation parameter. A variation in this parameter may be
compensated by the introduction of a constant noncommutative
magnetic
field. The usefulness of the noncommutative description will be
that
it might simplify the treatment of problems that are difficult to
deal
with in the usual commutative setting.

The organization of this paper is as follows: In
section~\ref{sec:ext}
we review some properties of a planar system of particles coupled
to a
strong magnetic field. In particular, we discuss the emergence of
area
preserving diffeomorphisms as symmetry transformations, and the
necessity of the introduction of a noncommutative geometry if a
consistent representation of the algebra of classical symmetries is
required. In particular, we discuss how the quantum version of
those
symmetries imply the noncommutativity of the gauge transformations.

In section~\ref{sec:eff}, the noncommutative description is
introduced
as a tool to change the part of the gauge field dynamics compatible
with those symmetries.  Finally, it is argued that a noncommutative
theory may be used as an effective description of the Quantum Hall
Effect, alternative to the usual CS commutative approach.

Some technical aspects of the path integral version of the Moyal
product, which are recalled in the main part of the article are
presented in Appendix A. Also, the apparently different way to
introduce the NCCS theory, based in the incompressible fluid
picture
is discussed in Appendix B.
 
\section{Matter current coupled to an external
field}\label{sec:ext}

In order to study the symmetries of the full system, as defined by
$S$
in equation (\ref{eq:defact}), it is useful to begin with the
simpler
case of a conserved matter current coupled only to an {\em
external\/}
gauge field.  The latter is assumed to correspond to a strong
uniform
magnetic field $B$, whose strength is supposed to be large when
compared with the interactions, in such a way that the dynamics can
be
safely restricted to the lowest Landau level. This assumption will
be
crucial in all our subsequent developments.

\subsection{Symmetries in the Lowest Landau level}

In a first quantized description, the form of the interaction term
in
the action, $S_{int}$, is supposed to be of the minimal type:
\begin{equation}
  \label{eq:defsint}
  S_{int}\;=\; \int d^3x \, A_{\mu}(x) j^\mu (x) \;,
\end{equation}
where $A_\mu$ denotes the gauge field corresponding to the purely
external (i.e., non dynamical) magnetic field. Later on we shall
also
include a fluctuating part $a_\mu$, so that in (\ref{eq:defsint})
we
will make the replacement: $A_\mu \to A_\mu + a_\mu$.  The most
important part
of the gauge field, determining the spectrum of the theory will be
assumed to be $A_\mu$, while $a_\mu$ is, from the time being,
assumed to
be perturbative in character.

To fix the ambiguity in the gauge field configuration corresponding
to
the constant magnetic field $B$, we will adopt the Weyl gauge
($A_0=0$), and a symmetric gauge choice for $A_j$:
\begin{equation}
  \label{eq:symmg}
  A_j({\vec x}) \;=\; - \frac{1}{2} \,B\,\epsilon_{jk} x^k \;.
\end{equation}
On the other hand, (\ref{eq:defsint}) also involves the spatial
part
of the matter field current, which for a system of $N$ particles
may
be written as:
\begin{equation}
  \label{eq:defcur}
  j^k(x^0,{\vec x}) \;=\; e\, \int dt \, \sum_{a=1}^N \,
\frac{dx^k_a(t)}{dt} \,
\delta(x^0-t)\, \delta^{(2)}({\vec x}-{\vec x}_a (t))   
\end{equation}
where $t \to {\vec x}_a (t), a=1,\ldots,N$ defines the particles'
trajectories.  Then,
\begin{equation}
  \label{eq:sint1}
S_{int} \,=\, e\, \sum_{a=1}^N \int dt \,A_k({\vec x}_a (t))\,
\frac{dx^k_a(t)}{dt} 
\;=\; \frac{b}{2} \, \int dt \, \sum_{a=1}^N \, {\dot x}^j_a(t)
\epsilon_{jk} x^k_a (t) 
\end{equation}
where $b \equiv e B$, and $k,j=1,2$. Using the first-order action
to define
the canonical momenta (or taking into account the second-class
constraints that follow from this first-order action) one sees that
the Poisson (Dirac) brackets are:
\begin{equation}
  \label{eq:dbs}
\{ x^j_a \,,\, x^k_b \}\;=\; \theta \, \delta_{ab} \, \epsilon^{jk}
\;\;,\;\;\;\; 
\theta \,=\, - b^{-1}\;\;,
\end{equation}
while for arbitrary functions $f$, $g$ of the coordinates one has
\begin{equation}
  \label{eq:dbsg}
\{f \,,\, g\}\;=\; \theta \, \sum_{a=1}^N  
\frac{\partial f}{\partial x^j_a}\, \epsilon^{jk} \, \frac{\partial
g}{\partial x^k_a}\;. 
\end{equation}
Since the theory is invariant under reparametrizations, the
canonical
Hamiltonian vanishes, and the theory is then also invariant under
the
full group of `canonical' transformations, namely, transformations
that leave the bracket (\ref{eq:dbs}) invariant. The infinitesimal
version of these transformations may be written as:
\begin{equation}
  \label{eq:cltrns}
\delta_\Lambda x^i \;=\; \eta \; \{ x^i , \Lambda(x)\}
\end{equation}
where $\eta$ is an infinitesimal constant, and $\Lambda(x)$ is an
arbitrary
function of the particles' coordinates ${\vec x}_a$.  These
transformations, and their finite counterparts, are symmetries of
the
classical action~\footnote{This is of course valid only if we
ignore
  $S_m$.}. Alternatively, they may be interpreted as time
independent
gauge transformations of the gauge field (the remaining gauge
freedom
in the $A_0=0$ gauge).  Indeed, under a standard (time independent)
gauge transformation, the Lagrangian changes in a total time
derivative:
\begin{equation}
  \label{eq:gtc}
\delta_\Lambda A_j (x) \;=\; \partial_j \Lambda (x)
\;\;\Rightarrow\;\;
\delta_\Lambda S_{int} \;=\; e \int dt \frac{d}{dt} \sum_{a=1}^N
\Lambda({\vec x}_a(t)) \;,
\end{equation}
implying that $\Lambda$ is the infinitesimal generator of the
canonical
transformations (\ref{eq:cltrns}) of the coordinates.

To understand the quantum realization of these symmetries, in the
canonical quantization approach, one imposes the fundamental
commutator
\begin{equation} 
 \label{eq:qcs}
[ {\hat x}^j_a \,,\, {\hat x}^k_b ] \;=\; i \,\hbar \,\theta
\,\delta_{ab} \,
\epsilon^{jk} \;,
\end{equation}
where we have written $\hbar$ explicitly, in order to trace the
quantum
effects.
 
To the classical canonical transformations there correspond the
quantum counterparts:
\begin{equation} 
 \label{eq:qcltrns}
{\hat x}^i \,\to\, {\hat x}^i_U \;=\; {\hat U}^\dagger {\hat x}_i
\hat U \;,
\end{equation}
where ${\hat U}$ is an arbitrary unitary operator. When the
transformation defined by ${\hat U}$ is connected to the identity,
the
infinitesimal version of (\ref{eq:qcltrns}) is, of course,
\begin{equation}
  \label{eq:qcltrns1}
{\hat x}^i \,\to\, {\hat x}^i + \delta_\Lambda {\hat x}^i \;\;,\;\;
\delta_\Lambda{\hat x}^i 
\;=\; \eta \; [ {\hat x}^i, {\hat \Lambda}] \;.
\end{equation}
To represent the classical symmetry generator by a quantum
operator,
one must adopt an ordering prescription. The product of operators
thus
ordered is not, however, compatible with the standard classical
composition rule for the product of functions. This is a well known
fact, usually presented in the context of the Weyl ordering
prescription~\cite{QG}.  Therefore, the noncommutativity in the
{\em
  classical\/} theory arises as a consequence of the usual operator
ordering problems of quantum mechanics, if one wants the mapping
between classical and quantum transformations to be consistently
defined.

One way to see this, is to write the Weyl-ordered operator
${\mathcal
  O}(f)$ associated to a classical function $f$ of the coordinates
in
a `Fourier' representation:
\begin{equation}
  \label{eq:w1}
{\hat {\mathcal O}}(f)\;=\; \int [\prod_{a=1}^N \frac{dp_1^a
dp_2^a}{2\pi \hbar}]\,
{\tilde f}(p) \, \exp [ \frac{i}{\hbar} \sum_{a=1}^N (p_1^a {\hat
x}^1_a +
p_2^a {\hat x}^2_a)]
\end{equation}
where
\begin{equation}
  \label{eq:w2}
{\tilde f}(p) \;=\; \int [\prod_{a=1}^N \frac{dx^1_a dx^2_a}{2 \pi
\hbar}]\,
f(x) \exp [- \frac{i}{\hbar} \sum_{a=1}^N (p_1^a x^1_a +p_2^a
x^2_a) ]
\end{equation}
is the Fourier transform of $f(x)$. Recalling that the Weyl-order
of a
product of operators is defined as the sum over all permutations of
the operators, it follows that (\ref{eq:w1}) is Weyl-ordered.
Notice
that the exponentials are Weyl-ordered (as can be checked by using
their series expansions), and that a linear combination of
Weyl-ordered operators is also Weyl-ordered. Thus, expression
(\ref{eq:w1}) may be thought of as a convenient way to
unambiguously
assign an operator ${\hat{\mathcal O}}(f)$ to a given classical
function of the coordinates, $f$. As already advanced, by bringing
the
product of two Weyl-ordered operators back to Weyl-order, one sees
that the resulting operator is not the one corresponding to the
usual,
commutative, product of the two classical functions, but rather to:
\begin{equation}
  \label{eq:w3}
{\hat {\mathcal O}}(f) \, {\hat {\mathcal O}}(g) \;=\; 
{\hat {\mathcal O}}(f \star g )   
\end{equation} 
where $\star$ denotes the Moyal product:
\begin{equation}
  \label{eq:w4}
f(x) \star g (x) \;=\; \exp ( \frac{i}{2}\hbar \,\theta
\epsilon^{jk}
\frac{\partial}{\partial \eta_j} \frac{\partial}{\partial\xi_k})
f(x+\eta) \; g(x+\xi) |_{\eta \to 0, \xi \to 0} \;.
\end{equation}

For infinitesimal transformations, we may use the expansion:
${\tilde
  f}(p)\,=\, 1 + \eta \, {\tilde \Lambda}(p)\, +\, {\mathcal
O}(\eta^2)$, so that
unitary operators may also be expanded as
\begin{equation}
  \label{eq:opexp}
{\hat U}(f)\;=\; {\hat I} \,+\, \eta \, {\hat T}_\Lambda \,+\,
{\mathcal O}(\eta^2)
\end{equation}
where
\begin{equation}
{\hat T}_\Lambda \;=\; \int [\prod_{a=1}^N \frac{dp_1^a
dp_2^a}{2\pi \hbar}]\,
{\tilde \Lambda}(p) \, 
\exp [ \frac{i}{\hbar} \sum_{a=1}^N (p_1^a {\hat x}^1_a +p_2^a
{\hat x}^2_a) ]
\;.
\end{equation}
Thus, for the composition of two infinitesimal transformations, the
change in ${\hat x}^j$ to the first order in each of the respective
infinitesimal parameters $\eta_{1,2}$ shall be given by
\begin{equation}
  \label{eq:compo}
\delta {\hat x}^j \;=\; \eta_1 \eta_2 \, 
[{\hat x}^j , [{\hat T}_{\Lambda_1} , {\hat T}_{\Lambda_2}]] \;.
\end{equation}
By (\ref{eq:w3}), we see that
\begin{equation}
   \label{eq:comm}
 [{\hat T}_{\Lambda_1} , {\hat T}_{\Lambda_2}] \;=\; {\hat
T}_{\Lambda_1 \star \Lambda_2
- \Lambda_2 \star \Lambda_1} \;.
\end{equation}
This shows that, in order to have a consistent unitary
representation
of the symmetry transformations, the Moyal commutator should
replace
the Poisson bracket in the classical theory. In particular, the
infinitesimal transformation of the coordinates is now
\begin{equation}
  \label{eq:nctrns}
\delta_\Lambda x^i_a \;=\;\frac{1}{i \hbar}\; \left( x^i_a \, \star
\, \Lambda \,-\, \Lambda \,\star\, x^i_a \right)
\end{equation}
which reduces to the Poisson bracket only in the $\hbar \theta \to
0$ limit.  The
combination $\hbar \theta=\frac{\hbar}{e B}$ can be interpreted as
the area per
particle that results from dividing the total area of the system by
the degeneracy of the Landau levels. The dimensionless combination
which may be used to give a meaning to the $\hbar \theta \to 0$
limit is the
ratio $\hbar \theta/l^2$, where $l$ is the typical scale of
variation of the
functions that appear in the Moyal bracket.  Therefore, if the
functions are smooth on the scale of $\hbar \theta$, the Moyal
product is
approximately the regular one.

We see then that the usual commutative product between functions of
the coordinates is replaced by the Moyal product. The latter
appears
naturally in the Weyl quantization prescription, and is a simple
reflection of the non-commutativity of the spatial coordinates.
The
existence of a minimal volume in the plane is, in this case, due to
the non-vanishing commutation relation between the coordinates,
which
play the role of conjugate variables (in the canonical sense).  The
cyclotron length sets the scale of the minimal area for this
problem.

Knowing what the transformation rules for the classical $x_i$
functions should be, it is clear that they do not correspond to a
standard gauge transformation of the gauge field. As we have
already
mentioned, a standard (time independent) gauge transformation
changes
the Lagrangian by a total time derivative, and therefore it is
equivalent to a canonical transformation of the coordinates
(\ref{eq:cltrns}).  On the other hand, the gauge field variation
corresponding to the transformations (\ref{eq:nctrns}) is:
\begin{equation}
  \label{eq:ncgt}
\delta_\Lambda  A_j(x) \,=\, \partial_j \Lambda (x)\,+\, \frac{1}{i
\hbar} \left( A_j (x) \star 
\Lambda (x) - \Lambda (x) \star A_j (x) \right) \;,   
\end{equation}
namely, they are $U(1)$ non-commutative gauge transformations.
These
are the gauge transformations we were looking for, and the gauge
field
action must, therefore, be constructed using this symmetry as a
criterion. It is worth remarking that this results agrees with the
somewhat different (but obviously related) approach
of~\cite{susskind}, if the full noncommutative version of the
latter
is used.

It is important to realize that the previous discussion on the
noncommutativity of the coordinates, and hence the `deformation' of
the ordinary product of classical functions into the Moyal product
is
independent of the gauge choice adopted for $A_j$. Indeed, had we
used
a gauge field in a general gauge (subject only to the condition
$\partial_1
A_2 - \partial_2 A_1 = B$) in the action $S_{int}$:
\begin{equation}
  \label{eq:gensint}
S_{int} \,=\, e\, \sum_{a=1}^N \int dt \,A_k({\vec x}_a (t))\,
\frac{dx^k_a(t)}{dt} 
\end{equation}
the canonical Poisson brackets would have been:
\begin{equation}
  \label{eq:gensint1}
  \{ x_a^k \,,\, e A_k (x_b) \} \;=\; \delta_{ab}  
\end{equation}
(no sum over $k$). Then, the use of the standard properties of the
Poisson bracket:
$$
\{ x^k_a \,,\, e A_k(x_b) \} \;=\; e \{ x^k_a \,,\,x^j_b
\}\,\partial_j
A_k(x_b)
$$
\begin{equation}
  \label{eq:gensint2}
=\;\frac{e}{2} \,\{ x^k_a \,,\,x^j_b \}\,(\partial_j A_k(x_b) -
\partial_k A_j(x_b)) 
\;=\; -\frac{1}{2\theta} \,\epsilon_{jk}\, \{ x^k_a \,,\,x^j_b \} 
\end{equation}
allows us to derive the same bracket as for the symmetric gauge
choice, namely,
\begin{equation}
  \label{eq:cpb}
\{ x^j_a \,,\, x^k_b \} \;=\; \theta \, \delta_{ab}\, \epsilon^{jk}
\;.
\end{equation}

\subsection{LLL projection and non-commutative description}

Let us now turn to the construction of the Hilbert space for the
one-particle first quantized system. This step is required to
implement the second quantization, since the one-particle states
are
indeed the building blocks of the Fock space. In canonical
quantization, one sees that the theory has, in Dirac's terminology,
two primary second-class constraints $\chi_1$, $\chi_2$:
\begin{equation}
  \label{eq:defcs}
\chi_1 \;=\; \pi_1 \,-\, e A_1(x) \;\approx\; 0 \;\;\;,\;\;\;
\chi_2 \;=\; \pi_2 \,-\, e A_2(x) \;\approx\; 0 \;.
\end{equation}
where $\pi_j=-i \partial_j$. Of course, there are many different
ways to
construct the quantum theory for this system, depending on the way
to
implement these constraints.  We have found it convenient to use an
approach which follows closely the physical situation corresponding
to
a non-relativistic particle of mass $m$ in the presence of an
external
magnetic field, when that magnetic field becomes very large. One
begins from the observation that $h$, the Hamiltonian for a single
particle of mass $m$ in a constant magnetic field $B$ may be
written
as
\begin{equation}
  \label{eq:cons1}
h\;=\; \frac{1}{2m} [(\pi_1 - e A_1)^2 + (\pi_2 - e A_2)^2 ]
\;=\; \frac{1}{2m} \, ( \chi_1^2 + \chi_2^2) \;.
\end{equation}
The constraints $\chi_1$ and $\chi_2$ are equivalent to the two
complex
combinations: $\chi = (\chi_1 - i \chi_2)/\sqrt{2}$, $\chi^* =
(\chi_1 + i
\chi_2)/\sqrt{2}$, which in the quantum theory become a pair of
mutually
adjoint operators:
\begin{equation}
  \label{eq:const2}
 {\hat \chi}\;=\;\frac{({\hat \chi}_1 - i {\hat \chi}_2)}{\sqrt{2}}
\;\;\;,\;\;\;
 {\hat \chi}^\dagger \;=\;\frac{({\hat \chi}_1 + i {\hat
\chi}_2)}{\sqrt{2}}
\end{equation}
verifying the commutation relation:
\begin{equation}
  \label{eq:const3}
[{\hat \chi}\,,\, {\hat \chi}^\dagger ] \;=\; \frac{\hbar}{\theta}
\;,
\end{equation}
which is independent of the gauge choice adopted for $A_j$.  These
two
second class constraints may also be thought of as a pair composed
by
a first-class constraint (${\hat \chi}$, say) plus its gauge fixing
(${\hat \chi}^\dagger$). This allows us to treat the constraints
differently,
by using an alternative interpretation. For example, one may just
use
Dirac's method for first class constraints, and demand the physical
subspace ${\mathcal H}_{phys}$ of the full Hilbert space ${\mathcal
  H}$ (i.e., the one constructed out of the unconstrained system)
to
be annihilated by the first class constraint
\begin{equation}
  \label{eq:phys}
{\hat \chi} \; | \psi \rangle \;=\; 0  \;\;\;\;\; \forall |\psi
\rangle 
\in {\mathcal H}_{phys} \;. 
\end{equation}

Thus, the definition of the physical Hilbert space can be
conveniently
defined as a `reduction' from the one corresponding to the usual
Hamiltonian for a particle in an external magnetic field. This
treatment of the constraints is of course the most convenient when
one
is indeed considering a physical situation described by the
Hamiltonian ${\hat h}$, since not only it describes the physical
Hilbert space (as a `vacuum'), but also it allows for the
consideration of the possible corrections due to the fact that the
reduction is a simplification of the real physical situation.
Indeed,
while the constrained manifold is defined by (\ref{eq:phys}),
corrections due to the kinetic term will be contained in higher
states, built upon the `vacuum' ${\mathcal H}_{phys}$.

To make this more explicit, one may introduce the operators:
\begin{equation}
  \label{eq:defaad}
{\hat a} \;=\; \sqrt{\frac{\theta}{\hbar}} \, {\hat \chi} \;\;\;
{\hat a}^\dagger  \;=\; \sqrt{\frac{\theta}{\hbar}} \, {\hat
\chi}^\dagger \;,
\end{equation}
which verify the standard creation and annihilation algebra,
\begin{equation}
  \label{eq:caa}
[ {\hat a}\,,\, {\hat a}^\dagger ] \;=\; 1 \;,
\end{equation}
while the Hamiltonian $h$ becomes:
\begin{equation}
  \label{eq:hada}
{\hat h} \;=\; \hbar \omega_c \, ( {\hat a}^\dagger {\hat a} \,+\,
\frac{1}{2} ) \;, 
\end{equation}
where $\omega_c \,=\,- \frac{e B}{m}\,=\,\frac{1}{m \theta}$ is the
cyclotron
frequency.  The lowest Landau level of the Hamiltonian ${\hat h}$
is
of course annihilated by ${\hat a}$, and the higher states may be
generated by repeated application of ${\hat a}^\dagger$: $|n\rangle
\,=\,
\frac{({\hat a}^\dagger)^n}{\sqrt n!} |0\rangle $.  All theses
states are,
however, degenerated. To treat this degeneracy one introduces the
operators ${\hat x}^1_0\,,\,{\hat x}^2_0$, which classically
correspond to the motion of the center of the trajectory and are
usually called guiding center coordinates. They are defined by:
\begin{eqnarray}
  \label{eq:def0}
{\hat x}^1_0 &=& {\hat x}^1 \,-\, \theta \, {\hat \chi}^2
\nonumber\\
{\hat x}^2_0 &=& {\hat x}^2 \,+\, \theta \, {\hat \chi}^1 \;,
\end{eqnarray} 
and verify the commutation relations:
\begin{equation}
  \label{eq:cr0}
[ {\hat x}^1_0 \,,\, {\hat x}^2_0 ] \;=\; i \hbar \theta \;, 
\end{equation}
rather than the usual commutativity, which holds between ${\hat
x}^1$
and ${\hat x}^2$:
\begin{equation}
  \label{eq:cr1}
[ {\hat x}^1 \,,\, {\hat x}^2 ] \;=\; 0 \;. 
\end{equation}
Besides, both ${\hat x}^1_0$ and ${\hat x}^2_0$ commute with ${\hat
  a}$ and ${\hat a}^\dagger$.

It is clear that the physical Hilbert space is the lowest Landau
level
of the Hamiltonian ${\hat h}$. Let us now consider how to define
physical operators, also in the Dirac approach. Being this a
first-class system, physical operators are to be defined as those
that
commute with the first class constraints, i.e., they are gauge
invariant. Thus what we need now is a procedure to assign a gauge
invariant operator to a given classical function of the
coordinates.
This `reduction' mechanism, and its relation to the Moyal product
is
now conveniently studied in terms of an arbitrary classical
function
$f(x)$ of the coordinates, and its corresponding operator
${\mathcal
  O}(f)$.  We begin by introducing a correspondence between
functions
and operators which is valid {\em before\/} reducing to the
physical
subspace, and then make the necessary changes. If $f(x)$ is
represented in terms of its Fourier transform in momentum space,
${\tilde f}(p)$:
\begin{equation}
  \label{eq:red1}
{\hat {\mathcal O}}(f)\;=\; \int \frac{dp_1 dp_2}{2\pi \hbar}\,
{\tilde f}(p) \, \exp [ \frac{i}{\hbar} (p_1 {\hat x}^1 + p_2 {\hat
x}^2)]
\end{equation}
then the product between classical functions is commutative, since
(\ref{eq:cr1}) implies that there are no ordering problems in the
definition of ${\hat{\mathcal O}}(f)$.  The noncommutativity arises
when writing ${\hat x}^i$ in terms of ${\hat x}^i_0$, so that
(\ref{eq:red1}) becomes:
\begin{equation}
  \label{eq:red2}
{\hat {\mathcal O}}(f)\;=\; \int \frac{dp_1 dp_2}{2\pi \hbar}\,
{\tilde f}(p) \, \exp [ \frac{i}{\hbar} (p_1 {\hat x}^1_0 + p_2
{\hat x}^2_0)]
\, \exp [ \alpha {\hat a}^\dagger - \alpha^* {\hat a} ]
\end{equation}
where ${\hat a}$ and ${\hat a}^\dagger$ are the operators defined
in
(\ref{eq:defaad}), and
\begin{equation}
  \label{eq:red3}
\alpha(p) \,=\,\frac{p_1 - i p_2}{\sqrt{2 m \hbar \omega_c}}
\;\;\;\;
\alpha^*(p) \,=\,\frac{p_1 + i p_2}{\sqrt{2 m \hbar \omega_c}} \;.
\end{equation}

It is obvious that, in general, a function so defined will not be
gauge invariant, since there are operators that do not commute with
the constraint (which is proportional to ${\hat a}$).  Indeed, the
gauge non-invariance of $f$ is due to the presence of the unitary
operator ${\hat D}(\alpha,\alpha^*)$, defined by
\begin{equation}
  \label{defu}
{\hat D}(\alpha,\alpha^*) \;=\; e^{\alpha {\hat a}^\dagger \,-\,
\alpha^* {\hat a}} 
\;,
\end{equation}
which produces shifts in $\alpha$ when acting on a coherent state
characterized by a complex number $\lambda$:
\begin{equation}
  \label{eq:actu}
{\hat D}(\alpha,\alpha^*) |\lambda\rangle \;=\; |\lambda + \alpha
\rangle \;\;,\;\;
{\hat a} |\lambda \rangle \;=\; \lambda |\lambda \rangle \;.
\end{equation}
On the other hand, ${\hat D}(\beta,\beta^*)$ is, indeed, the
unitary operator
that realizes the gauge transformations generated by the first
class
constraint, so that we may project ${\hat D}(\alpha,\alpha^*)$ into
its gauge
invariant part by taking the average with respect to the gauge
group:
\begin{equation}
  \label{eq:gruav}
{\hat D}_0 (\alpha,\alpha^*) \;=\;\frac{1}{\pi^2} \int d\beta
d\beta^* \,
{\hat D}^\dagger (\beta,\beta^*)\;{\hat D}(\alpha,\alpha^*)\;{\hat
D}(\beta,\beta^*)\;. 
\end{equation}
It is simple to check that:
\begin{equation}
  \label{eq:crule}
{\hat D}^\dagger (\beta,\beta^*)\;{\hat D}(\alpha,\alpha^*)\;{\hat
D}(\beta,\beta^*)
\;=\;{\hat D}(\alpha,\alpha^*) \, \exp[2 i {\rm Im}(\alpha
\beta^*)] 
\end{equation}
and this implies, after integrating over $\beta$ and $\beta^*$,
that:
\begin{equation}
  \label{eq:d0}
 {\hat D}_0 \;=\; 1
\end{equation}
where $1$ denotes the identity operator. Thus, we see that the
physical operator corresponding to $f$ is
\begin{equation}
  \label{eq:redf}
{\hat{\mathcal O}}_0(f) \;=\; \int \frac{dp^1 dp^2}{2 \pi \hbar} \,
{\tilde f}(p) \, \exp [ \frac{i}{\hbar} (p_1 {\hat x}^1_0 + p_2
{\hat x}^2_0)] 
\;,
\end{equation}
which, in view of the noncommutativity between the $x^j_0$
coordinates, will imply the Moyal product for the classical
functions.
It is important to realize that this reduction has been presented
here
entirely in terms of the constrained system, and not in the context
of
an approximation to the real situation where there are more levels
than just the vacuum. Had we wanted to keep the full Hilbert space,
then the projection would have to be understood as an operation
that
{\em changes the number of physical degrees of freedom}.  Still,
the
reduced operator could now be defined by taking the vacuum
expectation
value of (\ref{eq:red2}) on the lowest Landau level. This is a
partial
average, affecting only the annihilation and creation operators
that
go from one Landau level to the next one, leaving a dependence on
the
operators that take care of the degeneracy. Under this reduction in
the number of degrees of freedom, the operator ${\hat {\mathcal
    O}}(f)$ becomes ${\hat {\mathcal O}}_r(f)$, defined by
\begin{equation}
  \label{eq:red4}
  {\hat {\mathcal O}}_r(f)\;=\; {\mathcal N}\, \int \frac{dp_1
dp_2}{2\pi \hbar}\,
  {\tilde f}(p) \, \exp [ \frac{i}{\hbar} (p_1 {\hat x}^1_0 + p_2
{\hat x}^2_0)]
  \, \exp (-\frac{1}{2} |\alpha (p)|^2 ) \;,
\end{equation}
where ${\mathcal N}$ denotes a normalization constant, defined as
\begin{equation}
  \label{eq:defcaln}
{\mathcal N}^{-1} \;=\; \int \frac{dp_1 dp_2}{2\pi \hbar} \exp
(-\frac{1}{2} |\alpha|^2 )
\;,
\end{equation}
and introduced by reasons that will become clear later on.  Then
the
correspondence between functions and operators should be defined by
\begin{equation}
  \label{eq:red5}
  f(x) \;\to \; {\hat {\mathcal O}}_r(f)\;=\; \int \frac{dp_1
dp_2}{2\pi \hbar}\,
    {\tilde f}_r(p) \, \exp [ \frac{i}{\hbar} (p_1 {\hat x}^1_0 +
p_2 
{\hat x}^2_0)]
\end{equation}
where
\begin{equation}
  \label{eq:deff0}
{\tilde f}_r(p) \;=\; {\mathcal N} {\tilde f}(p) \exp (-\frac{1}{2}
|\alpha (p)|^2 ) 
\end{equation}
is a `smoothed' version of $f$. Indeed, in coordinate space, $f_r$
corresponds to $f$ convoluted with a Gaussian window of size equal
to
the cyclotron length for each coordinate. Of course, the Moyal
product
will now appear for the functions $f_r$, and not for the original
ones, $f$. This is to be expected, since the model with all the
Landau
levels as physical states is commutative, and some modifications
are
to be expected when comparing with the purely noncommutative model.
The normalization ${\mathcal N}$ is included in order to preserve
the
probability, when the reduction in the number of degrees of freedom
is
implemented.

Summarizing, we have shown that the proper treatment of the
constrained system naturally leads to the consideration, at the
classical level, of a non-commutative theory.  It should be noted
that
the original, commuting coordinates are mapped into the guiding
center
coordinates.

It is worth mentioning that everything we discussed here has its
analog formulation in the path integral quantization scheme, if the
proper translations are used. In particular, the Weyl ordering may
be
implemented by using the `mid-point prescription'.  The emergence
of a
noncommutative theory may also be shown to happen in the
path-integral
setting, as shown for point-splitting regularization in string
theory~\cite{SW}. This holds true also for the general case of
quantization deformation of a Poisson structure~\cite{cattaneo}. We
apply this to the case at hand in Appendix A, using the `magnetic'
language, and particularizing to the system of interest.

\section{Effective description for large magnetic fields}
\label{sec:eff}
 
As a description of a system with a large but finite magnetic
field, a
noncommutative formulation should, by the previous reasoning, be a
good approximation. However, it is unpleasant to realise that, in
fact, as soon as we assume that the gap between the lowest Landau
level and the upper ones is finite, the coordinates commute. This
discontinuous behaviour would seem to forbid any attempt to use the
noncommutative approach as a good starting point to deal with the
case
of a finite gap. The main reason for this discontinuous behaviour
is
of course that the number of physical degrees of freedom is
different
for the finite and infinite gap cases. In this sense the phenomenon
is
analogous to the CS quantum mechanics model of~\cite{jackiw}.  We
could attempt, however, an intermediate approach: the
noncommutative
theory could be introduced with a $\theta$ parameter corresponding
to a
strong magnetic field (not necessarily equal to the real external
one), but with the non-commutative theory still containing a
(noncommutative) external magnetic field. Indeed, for the single
particle action in an external field,
\begin{equation}
  \label{eq:sintg}
S_{int} \,=\, e\, \int dt \,A_k({\vec x}(t))\, \frac{dx^k(t)}{dt} 
\end{equation}
we may now assume that $A_k$ corresponds to a magnetic field $B$,
which can always be represented as
\begin{equation}
  \label{eq:decomp}
B \;=\; B^\theta \,+\, {\mathcal B} 
\end{equation}
where for some reason that depends on the physical problem one is
dealing with, $B^\theta$ is such that it results convenient to use
the
noncommutative description, and ${\mathcal B}\,=\,B-B^\theta$. Thus
the
idea is to go from the commutative description, where there is a
constant magnetic field $B$, to a noncommutative one with a
noncommutative parameter
\begin{equation}
  \label{eq:deftheta}
\theta \;=\; - \frac{1}{e B^\theta}
\end{equation}
and with a constant noncommutative magnetic field ${\hat {\mathcal
    B}}$, related to ${\mathcal B}$, as we shall see.  Indeed,
splitting also the gauge field $A$ (which verifies
$\vec{\nabla}\times {\vec A}
= B$), in two parts: $A^\theta$ and ${\mathcal A}$, such that
${\vec
  \nabla}\times{\vec A}^\theta = B^\theta$ and ${\vec
\nabla}\times{\vec {\mathcal A}} = {\mathcal
  B}$, we have the action describing the interaction:
\begin{equation}
  \label{eq:sintg1}
S_{int} \,=\, e\, \int dt \,A_k({\vec x}(t))\, \frac{dx^k(t)}{dt} 
=\,S^\theta \,+\, {\mathcal S} 
\end{equation}
where
\begin{equation}
S^\theta \,=\,e \, \int dt A^\theta_k({\vec x}(t))\,
\frac{dx^k(t)}{dt} 
\end{equation} 
and
\begin{equation}
{\mathcal S} \,=\, e \, \int dt{\mathcal A}_k({\vec x}(t))\, 
\frac{dx^k(t)}{dt}\;.  
\end{equation}
Then, the part of the action corresponding to $A^\theta$ is used to
introduce the noncommutativity, while the part proportional to
${\mathcal A}$, is treated as an external field for the remaining
theory.  However, in the noncommutative theory, this remaining
field
is not precisely equal to ${\mathcal A}$: When we consider gauge
transformations for ${\mathcal A}$, with $A^\theta$ fixed, the
classical
action is invariant, since these transformations change the
Lagrangian
by a total derivative. However, the quantum theory will not have
this
symmetry, by the same reason that made the action (\ref{eq:sint1})
invariant under (\ref{eq:ncgt}) rather than under the usual
commutative Abelian gauge transformations. Being a noncommutative
gauge field, we should write ${\hat \mathcal A}$ rather than
${\mathcal A}$ for the remaining gauge field in the noncommutative
theory, so that the action (\ref{eq:sintg1}) taking into account
quantum effects is now written as:
\begin{equation}
  \label{eq:sintg2}
{\hat S}_{int} \,=\,e \int \, dt {\hat A}_k({\vec x}(t))\,
\frac{dx^k(t)}{dt} 
\;.
\end{equation}
These quantum effects may be introduced by the device of using
$S^\theta
=e\int dt A^\theta_k \frac{dx^k}{dt}$ as the `free' action, which
then defines
the canonical structure and its associated Weyl ordering. A
possible
way to accomplish this can be to use the path integral framework to
derive the action ${\hat S}_{int}$ as the `effective' action that
results from a (partial) integration of the degrees of freedom,
namely,
\begin{equation}
  \label{eq:defeff}
e^{\frac{i}{\hbar}{\widehat S}_{int}[{\hat A}]}\;=\; 
\langle e^{\frac{i}{\hbar} {\mathcal S}}\rangle^\theta  
\end{equation}
where
\begin{equation}
\langle \cdots \rangle^\theta \;=\; \int {\mathcal D}x \, \cdots
e^{\frac{i}{\hbar} S^\theta} 
\end{equation}   
with the path integral evaluated in a semiclassical expansion,
defined
in the same way as in Appendix A. The resulting ${\widehat S}$
action
is of course noncommutative, since when expanding ${\mathcal S}$ in
(\ref{eq:defeff}), each product is replaced by its Moyal analog. We
note that also a perturbative field $a$ (not necessarily
corresponding
to a magnetic field but for instance to an external probe) will be
transformed into a noncommutative one by this device.

Of course, this procedure is not exact, since, had we used the full
gauge field as the free action, the canonical theory would have
been
different.  Besides, there would be no remnant field for this
different noncommutative theory, since in this case, we would have
traded all the magnetic field $B$ by $B_\theta$.  There is then an
interplay between the $\theta$ parameter and the strength of the
remaining
noncommutative field, which of course corresponds to a constant
noncommutative field strength $F_{ij}$.

To see this, we realize that to the usual $U(1)$ gauge orbits of
the
classical theory there will correspond gauge orbits of the
noncommutative $U(1)$ theory, so that the relation
\begin{equation}
  \label{eq:sw}
{\hat{\delta}}_{\hat{\lambda}}{\widehat{\mathcal A}} \;=\; 
\widehat{\delta_\lambda {\mathcal A}}
\end{equation}
which is the expression that leads to the Seiberg-Witten mapping
between commutative and noncommutative theories~\cite{SW}.  Thus,
if
not all the uniform magnetic field $B$ is traded by $B_\theta$ in
the
noncommutative description there is an extra constant
noncommutative
magnetic field.  To find a quantitative expression of this
interplay,
we may recall that a constant commutative magnetic field is mapped,
via the Seiberg-Witten equations, to a noncommutative constant
field
${\widehat{\mathcal B}}$, with the relation:
\begin{equation}
  \label{eq:sw1}
\frac{1}{e {\mathcal B}} \;=\; \frac{1}{e {\hat{\mathcal B}}} \,-\,
\theta 
\end{equation}
which is an exact solution of the SW relations, valid for the case
of
a constant magnetic field~\cite{SW}.

Equation (\ref{eq:sw1}) shows that if $\theta$ vanishes, the
noncommutative
description reduces to the usual commutative theory in a continuous
way. On the other hand , if all the magnetic field $B$ is traded by
$B_\theta$, there is no remaining magnetic field in the
noncommutative
theory. In this case, the limit of vanishing $\theta$ does not
reduces to
the original commutative theory anymore. This is of course
consistent
with the fact that for the Landau problem, the projection onto the
LLL
is not a continuous process since it implies a change in the
Hilbert
space of the system.

Based in the previously derived relations between the strong
magnetic
field Hamiltonian and a noncommutative theory, it should be noted
that
the classical action (to be used in second quantization) should
contain the Moyal product whenever products of functions of the
spatial coordinates appear. This is of course valid also for every
other term in the action, including a pair interaction term.  In
particular, it can be shown that an ultra-local pair interaction
term
in the noncommutative theory, can be mapped into the Hamiltonian
for a
free particle in a uniform magnetic field determined by $\theta$,
with an
effective mass proportional to the strength of the pair potential.
This term will play the role of an effective kinetic term for the
{\it
  projected\/} theory.

The standard second quantization action for a bidimensional system
of
non interacting particles in the presence of an external magnetic
field (before reducing to the lowest Landau level) would be:
$$
S_s = \int dt dx^1 dx^2 \psi^\dagger (t,x) \left[ i \hbar
\partial_t - e a_0 + \mu \right.
$$
\begin{equation}
  \label{eq:sqa}
\left. -\, \frac{1}{2m} (-i \hbar {\vec \nabla} - e {\vec A}_\theta
- e {\vec {\cal A}}
- e {\vec a})^2 \right]
 \psi (t,x) 
\end{equation}
where ${\vec A}_\theta$ and $\vec {\cal A}$ where defined above,
and $ a_\mu$
corresponds to an external probe.

Then the non-commutative description {\em with $B_\theta$
determining the
  noncommutativity\/} is introduced, as a reduction to the first
Landau level for $B_\theta$, passing from the action (\ref{eq:sqa})
to the
noncommutative one
$$
S_{nc} \;=\; \int dt dx^1 dx^2 \left[ \psi^\dagger (t,x)\star(i
\hbar \partial_t + \mu) \psi (t,x)
  - e \psi^\dagger(t,x)\star a_0 (t,x) \star \psi (t,x) \right.
$$
\begin{equation}
  \label{eq:sqa1}
\left. -\, \frac{1}{2m}\psi^\dagger(t,x) \star (-i \hbar {\vec
\nabla} - e 
( {\vec {\mathcal A}} + 
{\vec a}) )\star (-i \hbar {\vec \nabla} - e ({\vec \mathcal A} + 
{\vec a})) \right] \star \psi (t,x) 
\end{equation}
where the $B_\theta$ field part has disappeared from the action
(i.e., it
is in $\star$), since it has been traded for the noncommutativity
of the
coordinates:
\begin{equation}
\theta \;=\; - \frac{1}{e B_\theta} \;.
\end{equation}

An alternative way of justifying the introduction of the
noncommutative description in this context is as follows. We can
try
to decouple the matter fields from the uniform magnetic field by
performing a singular gauge transformation. In principle we can
write
\begin{eqnarray}
 \psi (t,x) &=& G_c (x) \psi_c(t,x) \nonumber \\
\psi^\dagger (t,x) &=&  \psi_c^\dagger (t,x) G_c^\dagger (x)
\end{eqnarray}
where
\begin{equation}
G_c(x)= \exp{{\frac{ie}{\hbar}}\int_{{\cal C}(x)} d{\vec
y}\cdot{\vec A}_{\theta}(y)}
\end{equation}
with ${\cal C}(x)$ denoting a curve that starts at spacial infinity
and ends at the point $\vec x$.  In this way the new fields are
free,
but at the cost of being dependent on the curve $\cal C$. However,
this dependence on the curve could be get rid off if, for any path
$\Gamma$, the condition
\begin{equation}
{{\frac{e}{\hbar}}\int_{\Gamma} d{\vec y} \cdot {\vec A}_\theta
(y)}= 
{\frac{e B_\theta}{\hbar}} \, S(\Gamma)= 2\pi n
\end{equation}
with $n \in \mathbb{Z}$ were satisfied. In this expression,
$S(\Gamma)$
denotes the area enclosed by the curve $\Gamma$. Thus, the gauge
transformation that eliminates the external magnetic field would be
independent of the path only if the area enclosed by an arbitrary
path
$\Gamma$ were quantized, i.e., if $S(\Gamma)= \hbar \theta 2\pi n$.
However, the
quantization of the area is difficult to justify, unless we work in
the context of non commutative geometry, were there is an
uncertainty
relation for the two spatial coordinates.  Notice that in the
Landau
problem the natural scale for the `quantum' of area is set by the
cyclotron length $l_0=\sqrt {\hbar \theta }$.

For a system of non-relativistic fermions in the presence of a
commutative gauge field with a part that corresponds to a uniform
magnetic field $B$ and a fluctuation $a_\mu$, the fermionic
determinant
can be calculated~\cite{lf} when the ratio between the average
density
and the magnetic field is such that there is an integer number of
Landau levels filled. In this case, the leading order term of the
effective action for $a_\mu$ has the Chern-Simons form, and its
coupling
constant is proportional to the ratio between the magnetic field
and
the average density (or the inverse of the filling fraction).

According to our previous discussion, we can apply the
Seiberg-Witten
transformation to this gauge field, with a $\theta$ parameter
defined by
$B_\theta$, so that the constant magnetic field $\mathcal B$ is
transformed
into ${\hat{\mathcal B}}$, through the relation (\ref{eq:sw1}).
Then
the commutative CS action is transformed into the noncommutative
one
\cite{gs} for the field ${\hat a}_\mu$ which is related to $a_\mu$
through
the Seiberg-Witten relation as well. We know that for vanishing
$\theta$
the CS action becomes the commutative one with a coupling constant
proportional to the inverse of the filling fraction. On the other
hand, if all the uniform magnetic field is traded by $B_\theta$,
there is
no induced Chern-Simons action. We will return to the problem of
the
coupling constant for an arbitrary $\theta$ elsewhere~\cite{IP}.

To finish this section we discus briefly a possible realization of
this approach in the context of the QHE problem.  It is well known
that in the presence of a strong perpendicular magnetic field, a
system of oppositely charged particles (such as a neutral dipole)
moves in a straight line perpendicular to the vector connecting
them,
even though its size grows with its momentum \cite{bs}. Such
dipoles
are the objects described by non-commutative field theories. In
particular, it has been shown that a set of local gauge invariant
operators in noncommutative gauge theories can be constructed by
using
straight Wilson lines with momentum $p_\mu$ such that the distance
between the end points of the line is $l^\nu = p_\mu
\theta^{\mu\nu}$
\cite{GHI,IIKK}. Given a local operator ${\cal O}(x)$ in an
ordinary
gauge theory (in the adjoint representation) its noncommutative
generalization is~\cite{GHI}
\begin{equation}
{\tilde{\cal O}}(k) = Tr \int d^3x {\cal O}(x) * P_* exp (iq \int_C
d\lambda^\mu 
A_\mu(x+\lambda) ) * e^{ikx}
\end{equation}
where $C$ is a straight path $\lambda^\mu (\sigma) = k_\mu
\theta^{\mu\nu} \sigma $, $0 \leq \sigma < 1$,
and $P_*$ denotes path ordering with respect to the star product.
The
tilde is used as a reminder that there is a Wilson line attached to
the operator.  The Wilson line is extended in the direction
perpendicular to the momentum. For small $k$ or $\theta$ the length
of the
Wilson line goes to zero and ${\tilde{\cal O}}$ reduces to the
corresponding operator in the commutative field theory.

In the context of the FQHE, it was argued in reference~\cite{read}
that for the half filled state, the true low-energy quasiparticles
in
the fermion Chern-Simons theory obtained upon screening of the
magentoplasmon mode, are electrically neutral (see also
\cite{read,ms,hp,dhl,hos}). Based on trial wave functions in the
LLL,
Read noticed that the electron and the correlation hole are
separated
from one another by a distance proportional and perpendicular to
the
canonical momentum $\vec k$ of these low energy quasiparticles.
Therefore, these neutral quasiparticles carry an electric dipole
moment $el^2 {\hat z} \times {\vec k}$ with $l$ the magnetic
length.  Thus,
if we choose the deformation parameter $\theta$ such that all the
external
magnetic field is traded by $B_\theta$ (i.e.  $B_\theta = B$), the
effective
theory (\ref{eq:sqa1}) (including a pair potential term, not
written
explicitly in that expression) will be an appropriate description
for
this problem, since it naturally describes the elementary
quasiparticles of the half-filled state.  There are a couple of
results that support our proposal. In a similar model studied in
reference \cite{wf} the authors show that the corresponding ground
state wave function has the shifting between the particle and the
correlation hole discussed by Read \cite{read}. We also know
\cite{IP}
that this model breaks parity without the presence of an explicit
Chern-Simons term, in coincidence with the description of reference
\cite{read}.


\section{Conclusions}\label{sec:con}

In this work we have studied different aspects of the description
of
two dimensional systems in high magnetic fields using
noncommutative
theories.  We began by reviewing the problem of a particle coupled
to
a magnetic field whose magnitude is large enough to neglect the
kinetic energy.  In this case, the spatial coordinates are
canonical
conjugate to each other, and the system is invariant under area
preserving diffeomorphisms of the plane. Thus, at the quantum level
the Moyal bracket should replace the Poisson bracket for
infinitesimal
coordinate transformations. Alternatively, one may think in terms
of
gauge transformations for the gauge field coupled to the particle.
In
this case, the usual gauge transformations are replaced by their
noncommutative version. Therefore, the gauge field action must be
constructed being invariant under this noncommutative gauge
symmetries.  In references \cite{FZ,FK} it was argued using general
hydrodynamical arguments that the effective action for an
incompressible state of a system of charged particles in two
dimensions in the presence of a strong magnetic field must be a
Chern-Simons action.  Analogously, and using the fact that the {\it
  correct\/} symmetry for the gauge fields in the LLL is the
noncommutative gauge symmetry, the {\it natural\/} effective
description should be given by the noncommutative Chern-Simons
action.

As we have already mentioned, the Hilbert space is not the same in
the
case that all the Landau levels are taken into account, than if
only
the LLL can be occupied.  In particular, the space coordinates
commute
in the first case, and they do not in the latter, and the number of
degrees of freedom is different. In this sense the {\it
projection\/}
onto the LLL can not be made in a continuous way.  We have argued
that
if the only allowed state is the LLL, the correct description is a
noncommutative free theory. Therefore the obvious question is how
to
make compatible the noncommutative description with a problem in
which
some Landau level mixing is present.  We argued that in this case
it
should be used a someway intermediate approach. The noncommutative
theory could be introduced with a $\theta$ parameter corresponding
to a
magnetic field $B_\theta$ (not necessarily equal to the external
one), but
with the non-commutative theory still containing a (noncommutative)
uniform magnetic field $\hat {\cal B}$, in such a way that $\hat
{\mathcal B}$ is related to $\mathcal B$ through the Seiberg-Witten
relation, and the external uniform magnetic field is $B=B_\theta +
\hat
{\mathcal B}$.  Then we argued that once the fermionic determinant
is
calculated for this theory, the leading order term in a derivative
expansion will be given by a NCCS action for the external probe
whose
coupling constant will be a function of $\theta$ and $\hat
{\mathcal B}$.

To conclude, we mention that noncommutative field theories have an
unusual perturbative behaviour. This is due to the fact that the
Moyal
product generates phases appearing in the perturbative structure
that
induce an interplay between the infrared and the ultraviolet
regimes.
It can be argued that since spacial non-commutativity is a short
distance property, it would be surprising that some effect related
to
it could show up in the low energy effective theory. However, it
was
shown that for some non-commutative field theories
\cite{MST,cri1,cri2} the noncommutativity of the coordinates
modifies
the critical behaviour of the theory, since the long distance
behavior
is entangled to the short distance one due to the presence of the
Moyal phases. This interplay between short and long distance
behaviour
therefore changes the critical properties of the noncommutative
theories compared to their commutative counterparts. In this
context,
we believe it might prove useful to explore the alternative {\it
  noncommutative descriptions\/} of bidimensional systems in high
magnetic fields described in this work, to approach problems where
their commutative counterparts fail.

\section*{Acknowledgments}
We thank Prof. M. Aurelio for enlightening discussions.  This work
is
supported by CONICET (Argentina), by ANPCyT through grant No.\ 
$03-03924$ (AL), and by Fundaci{\'o}n Antorchas (Argentina).

\section*{Appendix A: Path integral representation of 
  the Moyal product} The Moyal product of two functions of the
coordinates $f, g$ may be represented, following~\cite{cattaneo},
in
terms of a quantum mechanical path integral with a topological
action.
This action becomes particularly simple when one considers the
deformation quantization of a Poisson structure defined by a
symplectic form, and this is, indeed, the case at hand.

For this simple case, the expression for the Moyal product may be
written as
\begin{equation}
   \label{eq:frep}
(f \star g)(x)\;=\; \int_{\gamma(\pm \infty)=x} {\mathcal D}\gamma
\; f(\gamma(1)) g(\gamma(0)) \; e^{\frac{i}{\hbar}S[\gamma]} 
\end{equation}
where $\gamma:{\mathbb R}\to\mathbb{R}^2$ denotes a plane curve,
and the
action $S[\gamma]$ is defined by
\begin{equation}
  \label{eq:frep1}
S[\gamma]\;=\; \frac{b}{2} \, \int_{-\infty}^{+\infty} dt \; {\dot
\gamma}^j(t) \epsilon_{jk} \gamma^k(t) \;, 
\end{equation}
with $b \,=\,-\,\theta^{-1}$. It is also adopted as a prescription
that the
functional integral should be evaluated semiclassically, around the
`classical' configuration \mbox{$\gamma^j (t) = x^j = {\rm
constant}$}.
This path integral formula may also be thought of as a concrete
realization of Kontsevich's result on the expression of the star
product in a Feynman-like perturbation expansion~\cite{kontsevich}.

In the case at hand, the above definition may be applied to two
functions $f$ and $g$ more directly if they are written in terms of
their Fourier transforms:
\begin{eqnarray}
f(x)&=& \int \frac{d^2k}{(2\pi)^2}\, {\tilde f}(k) \, 
e^{i {\vec k} \cdot {\vec x}}\nonumber\\ 
g(x)&=& \int \frac{d^2k}{(2\pi)^2}\, {\tilde g}(k) \, 
e^{i {\vec k} \cdot {\vec x}} 
\end{eqnarray}
so that (\ref{eq:frep}) becomes
$$
\int_{\gamma(\pm \infty)=x} {\mathcal D}\gamma \, f(\gamma(1))
g(\gamma(0)) \,
e^{\frac{i}{\hbar}S[\gamma]} \;=\; \int \frac{d^2k}{(2\pi)^2}
\frac{d^2l}{(2\pi)^2}
$$
\begin{equation}
  \label{eq:frep2}
\times {\tilde f}(k) {\tilde g}(l) \, \int_{\gamma(\pm \infty)=x}
{\mathcal D}\gamma \, 
\exp\{\frac{i}{\hbar} S[\gamma] + i \int_{-\infty}^{+\infty} dt
\,\gamma^j(t)\,[k^j \delta(t-1) +l^j \delta(t)]\}\;,
\end{equation}
where the plane wave parts of the Fourier transforms have been
included in the source term of ${\gamma}^j(t)$.  We then make a
shift in
the integration variables: $\gamma^j(t) \to x^j + \xi^j(t)$, so
that the
measure is now ${\mathcal D}\xi$, and $\xi$ vanishes at
$\pm\infty$:
$$
\int_{\gamma(\pm \infty)=x} {\mathcal D}\gamma \, f(\gamma(1))
g(\gamma(0)) \,
e^{\frac{i}{\hbar}S[\gamma]} \;=\; \int \frac{d^2k}{(2\pi)^2}
\frac{d^2l}{(2\pi)^2} \,
{\tilde f}(k) {\tilde g}(l) \, e^{i ({\vec k}\cdot {\vec x} + {\vec
l}\cdot
  {\vec x})}
$$
\begin{equation}
  \label{eq:frep3}
\times \int_{\xi (\pm \infty)=0} {\mathcal D}\xi \,
\exp\{\frac{i}{\hbar} S[\xi] + i \int_{-\infty}^{+\infty} dt
\,\xi^j(t) \,
[k^j \delta(t-1) + l^j \delta(t)]\} \;.
\end{equation}
Thus the integral over $\xi$ is a Gaussian and we may write its
result
explicitly:
$$
\int_{\xi (\pm \infty)=0} {\mathcal D}\xi \, \exp\{\frac{i}{\hbar}
S[\xi] + i
\int_{-\infty}^{+\infty} dt \xi^j(t) [k^j \delta(t-1) +l^j
\delta(t)]\}
$$
\begin{equation}
  \label{eq:frep4}
=\,\exp\{-\frac{i \hbar}{2 b} \int dt_1 \int dt_2 [k^i
\delta(t_1-1)
+l^i \delta(t_1)]K_{ij}(t_1-t_2) [k^j \delta(t_2-1) +l^j
\delta(t_2)]\}
\end{equation}
where $K_{ij}(t)$ is the inverse of the operator defining the
quadratic form in the action, namely,
\begin{equation}
  \label{eq:frep5}
- \epsilon^{ij} \frac{d}{dt} K_{jk}(t) \;=\; \delta (t) \delta^i_j 
\end{equation}
which has the solution
\begin{equation}
  \label{eq:frep6}
K_{ij} (t) \;=\; \frac{1}{2} \epsilon_{ij} {\rm sign}(t) \;.
\end{equation}
This propagator is uniquely defined, since it has to be Bose
symmetric: $K_{ij}(t)=K_{ji}(-t)$, and moreover it is also
consistent
with the canonical commutator of Equation (\ref{eq:qcs}). The last
condition may be verified by a direct application of the BJL
limit~\cite{BJL} to derive the equal time commutator between
$\gamma^j$ and
$\gamma^k$:
\begin{equation}
  \label{eq:bjl}
[\,\gamma^j(t)\,,\,\gamma^k(t)\,] \;=\; (\lim_{\varepsilon \to
0^+}- \lim_{\varepsilon \to 0^-}) \langle \gamma^j(t+\varepsilon)
\gamma^k(t) \rangle  
\end{equation} 
where $\langle \gamma^j(t_1) \gamma^k(t_2) \rangle$ is the
propagator derived from
(\ref{eq:frep}).  This is of course proportional to the inverse of
$K_{ij}$:
\begin{equation}
  \label{eq:bjl1}
\langle \gamma^j(t_1) \gamma^k(t_2) \rangle \;=\; i \frac{\hbar}{2
b} \,\epsilon^{jk}\, {\rm sign}(t_1-t_2) \;,
\end{equation}
and when inserted in (\ref{eq:bjl}) reproduces the commutator we
had
obtained by canonical means in (\ref{eq:qcs}) for the coordinates
of
the particles in an external field.

Using now the explicit form for $K$ in (\ref{eq:frep4}), we find
\begin{equation}
   \label{eq:frep7}
\int_{\xi (\pm \infty)=0} {\mathcal D}\xi \, \exp\{\frac{i}{\hbar}
S[\xi] + i \int_{-\infty}^{+\infty} dt \xi^j(t) [k^j \delta(t-1)
+l^j \delta(t)]\}= e^{-\frac{i \hbar}{2 b}\epsilon_{ij} k^i l^j }
\;.
\end{equation}
which inserted in (\ref{eq:frep3}) yields
\begin{equation}
\label{eq:frep8} 
 \int_{\gamma(\pm \infty)=x} {\mathcal D}\gamma \, f(\gamma(1))
g(\gamma(0))\, e^{\frac{i}{\hbar}S[\gamma]}\,=\,
\left[\exp(\frac{i \hbar \theta}{2} \frac{\partial}{\partial x^j}
\frac{\partial}{\partial y^k} ) f(x) g(y)
\right]_{y \to x}   
\end{equation}
where the last expression is of course one of the possible ways to
define the Moyal product $(f \star g)(x)$, with a parameter $\theta
= -
b^{-1}$.

\section*{Appendix B: Fluid representation}\label{sec:fluid}
We discuss here some aspects of a somewhat different approach to
the
introduction of a noncommutative CS theory, this time in terms of
of a
fluid representation. Although this is not the path to the NCCS
theory
that we have followed in the main part of this article, we have
nevertheless included it here, for the sake of completeness.
Besides,
we consider here a different version of the approach developed in
ref.\cite{susskind}, which is applicable to the more general case
of
bosonized theories in $2+1$ dimensions, supplemented by an
incompressibility constraint.  Our starting point is the expression
for the bosonized action $S_B[A]$, which in the leading
approximation
in a derivative expansion is given by
\begin{equation}
  \label{eq:boson}
S_B[A] \;=\; S_{CS}[A] 
\end{equation}
where $S_{CS}$ denotes the CS action:
\begin{equation}
  \label{eq:defscs}
S_{CS}[A] \;=\; \frac{\kappa}{2} \int d^3x
\,\epsilon_{\mu\nu\lambda} 
A_\mu \partial_\nu A_\lambda \;. 
\end{equation}
This gauge field is related to the vacuum expectation value (VEV)
of
the bosonized matter current by
\begin{equation}
  \label{eq:br}
\langle J^\mu(x) \rangle \;=\; j^\mu (x) \;=\; 
\epsilon^{\mu\nu\lambda} \partial_\nu A_\lambda (x) \;.
\end{equation}
In the $A_0=0$ gauge, the spatial components of the current are
\begin{equation}
  \label{eq:br1}
j^k (t,{\vec x}) \;=\; - \epsilon^{kl} \frac{\partial}{\partial t}
\, 
A_l (t,{\vec x})\;.
\end{equation}
To go to the fluid interpretation, one regards the spatial current
as
a density $\rho$ times the fluid's velocity ${\vec v}$. From this
expression, we may formally write the equation that determines the
fluid flux lines:
\begin{equation}
  \label{eq:fl}
\frac{\partial x^k}{\partial t} \;=\; - \frac{1}{\rho} \,
\epsilon^{kl} \frac{\partial A_l}{\partial t}          
\end{equation}
and its solution shall be of the form:
\begin{equation}
  \label{eq:fls}
x^k \;=\; x^k(t,{\vec y})
\end{equation}
where ${\vec y}$ denotes the initial conditions for a given line.
Namely, a given value of ${\vec y}$ determines one line from its
initial point at $t=t_0$. Note that, in principle, both $\rho$ and
$A$
may be functions of the space and time coordinates. In order to
proceed from equation (\ref{eq:fl}), we need to make further use of
the continuity equation, and introduce the area preserving
diffeomorphisms symmetry assumption.  A convenient way to do this
is
by defining a 2-form $\Omega$ by
\begin{equation}
  \label{defom}
\Omega \;=\; \rho (dx^1-v^1 dt) \land (dx^2-v^2 dt) 
\end{equation}
which, by some elementary algebraic steps can be shown to verify:
\begin{equation}
  \label{eq:defo1}
d \Omega \;=\; [ \frac{\partial}{\partial t} \rho \,+\, {\vec
\nabla}\cdot{\vec j} ] \, dt \land dx^1 \land dx^2 \;=\; 0\;,
\end{equation}
as a consequence of the continuity equation (an assumption of the
bosonization approach). When writing $\Omega$ in terms of the
formal
solutions (\ref{eq:fls}), one makes use of $dx^i = \frac{\partial
x^i}{\partial t}
dt + \frac{\partial x^i}{\partial y^j} dy^j$ and $v^i =
\frac{\partial x^i}{\partial t}$ to
obtain:
\begin{equation}
  \label{eq:defo2}
\Omega \;=\; \rho \frac{\partial (x^1,x^2)}{\partial (y^1,y^2)} \,
dy^1 \land dy^2 \;.
\end{equation}
Equation (\ref{eq:defo1}) holds for any choice of coordinates, and
in
this set implies:
\begin{equation}
  \label{equation}
\frac{\partial}{\partial t} [\rho \frac{\partial
(x^1,x^2)}{\partial (y^1,y^2)}] \;=\; 0\;.  
\end{equation}
We now impose the area preservation requirement to the system,
namely,
\begin{equation}
  \label{eq:apt}
 \frac{\partial (x^1,x^2)}{\partial (y^1,y^2)} \;=\; 1
\end{equation}
so that $\rho = \rho_0$. Choosing the ${\vec y}$ coordinates in
order to
have a constant $\rho_0$, we have a uniform and constant density.
With
this in mind, (\ref{eq:fl}) can be integrated, yielding
\begin{equation}
  \label{eq:flsol}
x^k (t,{\vec y}) \;=\; y^k \,-\, \frac{1}{\rho_0} \, \epsilon^{kl}
A_l (t,{\vec y}) \;,
\end{equation}  
which is, indeed, the relation introduced in ref(). Up to now, we
have
used just the bosonization rule that yields the VEV of the current
in
terms of the curl of the gauge field $A_\mu$, without actually
using the
explicit form of the bosonized action. It turns out that the
invariance under area preserving diffeomorphisms is not compatible
with the standard Chern-Simons action. This may be seen from the
relation (\ref{eq:flsol}), which, when applied to the area element,
yields
\begin{equation}
  \label{eq:apt1}
dx^1 \land dx^2 \;=\; [ 1 - \frac{1}{\rho_0} {\mathcal B} ] dy^1
\land dy^2
\;=\; dy^1 \land dy^2  
\end{equation}
where
\begin{equation}
  \label{eq:defcb}
{\mathcal B} \;=\; \partial_1 A_2 - \partial_2 A_1 -
\frac{1}{\rho_0} \{ A_1 , A_2 \} 
\end{equation}
with $\{A,B\}= \epsilon^{jk} \partial_j A \partial_k B$. Thus, one
must impose the constraint
${\mathcal B}=0$.  As discussed in ref\cite{susskind}, this
constraint, together with the `kinetic' term for the fluid may be
written in a way which is tantamount to the first non-trivial
approximation to the noncommutative Chern-Simons action. Thus we
may
certainly conclude that the bosonization mapping between the
current
and the gauge field leads naturally to a fluid interpretation, and
that this fluid may be described by a non linear Chern-Simons like
action which is an approximation to the full noncommutative theory.
However, it is easy to see that, even in the context of this
approximate bosonization, the noncommutativity is bound to arise
when
including quantum effects.  Indeed, one way to see this is from the
fact that the fluid coordinates $x^i$ will be correlated by quantum
(loop) effects.  Since the coordinates are proportional to the
components of $A$, the existence of a nontrivial correlation
between
the two different components of $A$'s in the quantum theory will be
translated into a non trivial correlation for the corresponding
coordinates.  By the BJL limit, this correlation implies the
noncommutativity of the coordinates in the quantum version of the
theory, and hence the noncommutativity of the CS action.  The
correlation of the $A$'s, on the other hand, is due to the fermion
loop, and in this approximation is given by the (commutative)
Chern-Simons action.

\newpage


\begin{thebibliography}{bib}
  
\bibitem{castellani}L.~Castellani, Class.~Quant.~Grav.~17 (2000)
3377
  [hep-th/0005210], and references therein.
  
\bibitem{DN}M.~Douglas and N.~Nekrasov, hep-th/0106048, and
references
  therein.
  
\bibitem{susskind}L.~Susskind, hep-th/0101029.
  
\bibitem{PP}A.~P.~Polychronakos, JHEP {\bf 0104} 011 (2001)
  [hep-th/0103013].
  
\bibitem{KS}D.~Karabali and B.~Sakita, hep-th/0106016
  
\bibitem{PA}V.~Pasquier, Phys.~Lett.~{\bf B 490} (2000) 258;
  V.~Pasquier, cond-mat/0012207.
  
\bibitem{LMR}B.~H.~Lee, K.~Moon, and C.~Rim, hep-th/0105127.
skyrmions.
  
\bibitem{IK}I.~I.~Kogan, Int.~J.~Mod.~Phys.~A9 (1994) 3887.
  
\bibitem{SS}B.~Sakita, Phys.~Lett.~B315 (1993) 124;
Interactions 
  Rashmi Ray and B. Sakita, Ann.~Phys. 230 (1994) 131.
  
\bibitem{GCT}A.~Capelli, C.~Trugenberger and G.~Zemba, Nucl. Phys
{\bf
    B396} (1993) 465.
  
\bibitem{gj}S.~M.~Girvin, and T.~Jach, Phys.~Rev. {\bf B 29} (1984)
  5617.
  
\bibitem{jackiw}G.~V.~Dunne, R.~Jackiw, and C.~A.~Trugenberger.
Phys.
  Rev {\bf D41} (1990) 661.
  
\bibitem{midpoint}W.~Greiner, and J~Reinhardt, ``Field
Quantization'',
  Springer-Verlag (1996).  \bibitem{QG} V.~Chari and A.~Presley,
``A
  guide to Quantum Groups'', Cambridge University Press (1994).
  
\bibitem{SW}N.~Seiberg, and E.~Witten, JHEP 09 (1999) 032.
  [hep-th/9908142]
     
\bibitem{cattaneo}A.~S.~Cattaneo and G.~Felder,
formula,''
  Commun.\ Math.\ Phys.\ {\bf 212} (2000) 591 [math.qa/9902090].
  
\bibitem{lf}A.~Lopez and E.~Fradkin, Phys. Rev. {\bf B44}, 5246
  (1991).
  
\bibitem{gs}N.~Grandi and G.~A.~Silva, hep-th/0010113.
  
\bibitem{bs}D.~Bigatti and L.~Susskind, Phys. Rev. {\bf D62} 066004
  (2000).
  
\bibitem{GHI}D.~J.~Gross, A.~Hashimoto and N.~Itzhaki,
hep-th/0008075.
  
\bibitem{IIKK}N.~Ishibashi, S.~Iso, H.~Kawai and Y.~Kitazawa, Nucl.
  Phys. {\bf B573}, 573 (2000).
  
\bibitem{read}N.~Read, Semi.~Cond.~Sci.~Tech. {\bf 9}, 1859 (1994);
  Surf. Sci. {\bf 361/362} 7 (1996), Phys.~Rev.~{\bf B 58} (1998)
  16262.
  
\bibitem{ms}G.~Murthy and R.~Shankar in ``Composite Fermions'',
ed.\ 
  by O.~Heinonen, (North Holland, 1998).
  
\bibitem{hp}V.~Pasquier, and F.~D.~M.~Haldane, Nucl.~Phys.  {\bf
B516}
  719 (1998).
  
\bibitem{dhl}D.~H.~Lee, Phys.~Rev.~Lett {\bf 80} 4547 (1998).
  
\bibitem{hos}A.~Stern, B.~I.~Halperin, F.~von Oppen, and S.~Simon,
  Phys. Rev. {\bf B 59} (1999) 12547.
  
\bibitem{wf}D.~Bak, S.~K.~Kim, K.~S.~Soh, and J.~H.~Yee, Phys. Rev.
  Lett. {\bf 85}, 3087, (2000).
  
\bibitem{IP}C.~D.~Fosco, and A.~Lopez, unpublished.
  
\bibitem{FZ}J.~Frolich, and A.~Zee, Nucl.~Phys. {\bf B364} 517
(1991).
  
\bibitem{FK}J.~Frolich, and T.~Kerler, Nucl.~Phys. {\bf B354} 369
  (1991).
  
\bibitem{MST}A.~Matusis, L.~Susskind, N.~Toumbas, JHEP 0012 (2000)
002
  
\bibitem{cri1}J.~Gomis, K.~Landsteiner, and E.~Lopez,
hep-th/0004115.
  
\bibitem{cri2}G.~H.~Chen, and Y.~S~Wu, hep-th/0103020.
  
\bibitem{kontsevich}M.~Kontsevich, ``Deformation quantization of
  Poisson manifolds, I,'' q-alg/9709040.
  
\bibitem{BJL}J.~D.~Bjorken, Phys. Rev. {\bf 148}, 1467 (1966);
  K.~Johnson and F.~E.~Low, Prog. Theoret. Phys.(Kyoto), Suppl.{\bf
    37-38}, 74 (1966).

\end{thebibliography}
\end{document}